\documentclass[%
reprint,
 amsmath,amssymb,
 aps,
prb,
]{revtex4-2}

\usepackage{ulem}
\usepackage{xcolor}
\usepackage{soul}
\usepackage{graphicx}
\usepackage{dcolumn}
\usepackage{bm}
\usepackage{float}
\usepackage{hyperref}
\usepackage{todonotes}
\usepackage{ulem}

\begin{document}

\preprint{APS/123-QED}

\title{Theory and experiments of spiral unpinning in the Belousov-Zhabotinsky reaction using a circularly polarized electric field }

\author{Amrutha S V}
\thanks{Copyrights Reserved}%
\author{Anupama Sebastian}%
\author{Puthiyapurayil Sibeesh}
\author{Shreyas Punacha}
\author{T K Shajahan}
\email{shajahantk@nitk.edu.in}
\affiliation{
 Department of Physics\\ National Institute of Technology Karnataka}

\date{\today}

\begin{abstract}

We present the first experimental study of unpinning a spiral wave of excitation using a circularly polarized electric field. The experiments are conducted in the Belousov-Zhabotinsky(BZ) reaction, and the system is modeled using the Oregenator model. The mechanism of unpinning in the BZ reaction differs from that in the physiological medium. 
We show that the wave unpins when the electric force opposes
the propagation of the spiral wave. We developed an analytical relation of the unpinning phase
with the initial phase, the pacing ratio, and the field strength and verified the same.

\end{abstract}

                             
\maketitle

The Belousov-Zhabotinsky (BZ) reaction has served as the prototype of a large
class of systems that display excitation waves, including the waves of action
potentials seen in the heart~\cite{Luther2011}, brain~\cite{Huang2004}, retina~\cite{yu2012reentrant},
and waves of communication in the social amoeba dictyostelium discoideum~\cite{noorbakhsh2015modeling,kamino2017fold}.  
Excitation waves in these systems exhibit strikingly similar spatio-temporal 
patterns such as expanding target waves or rotating spiral
waves~\cite{zhang2019stability,zykov2018spiral,sinha2014patterns}.  Recently there
has been a renewed interest in the pattern formation in the BZ reaction because of the
active nature of the chemical waves: their wavefronts are electrically
charged~\cite{Amrutha,steinbock1992electric} and resultant changes 
in the surface tension on the droplets of BZ reagents in an oily medium
can propel the droplets~\cite{jin2017chemotaxis, ryabchun2022run}.

A characteristic feature of excitation waves is their
tendency to pin to heterogeneities in the
medium~\cite{sutthiopad2015propagation, lim2006spiral,Zemlin2012, Shajahan:07}. A 
pinned rotating wave requires a carefully administered stimulus to remove it from the heterogeneity~\cite{Bittihn2010}.
This is especially
pertinent in cardiac tissue since stable pinned rotating waves can be
life-threatening~\cite{Gray2011, Bittihn2010}.

Several groups have proposed methods for controlling such pinned waves using either
pulsed electric field~\cite{Luther2011,punacha2019spiral} or, more recently, circularly
polarized electric field~\cite{Feng2014, pan2016removal,feng2022removal}. 
Numerical studies have shown that circularly polarized electric fields (CPEF) are
more efficient in controlling cardiac excitation
waves~\cite{feng2022removal,punacha2020theory,Feng2014}. 
In particular, CPEF requires less energy and is more efficient in 
controlling pinned rotating waves~\cite{Feng2014, pan2016removal}. Our systematic investigations on the mechanism of CPEF 
indicated that the spiral wave could be unpinned if the frequency of the CPEF is more than a cut-off frequency~\cite{punacha2020theory}. 

It is observed that chemical waves are also prone to pinning~\cite{sutthiopad2015propagation}, and they can
also be unpinned using electric field~\cite{Amrutha,sutthiopad2014unpinning}. However, there is an essential distinction between the chemical wave and the waves in physiological
tissue. In the latter, the electric force does not affect the excitation wave
directly, instead, they unpin by inducing secondary excitations from the
heterogeneities~\cite{pumir2007wave}. In the chemical medium, on the other hand,
the wavefront
contains charged ions such as $Br^-$ and $Fe^{3+}$, which can be moved by the applied
electric field, {\it, i.e.,} the electric field in a
chemical medium exerts an advective force directly on the wavefront~\cite{Amrutha,jimenez2013electric,agladze1992influence}. 
Such an electric force on the wave is not reported in the physiological tissue. It is
also observed that the chemical wave unpins  
as it moves away from the anode, and not when moving towards it~\cite{Amrutha}.  

So far, there have not been any experimental reports of unpinning spiral waves using CPEF, either in the chemical
medium or the cardiac tissue. However, CPEF is realized in BZ medium to control spiral turbulence~\cite{ji2013experimental}. In this paper, we report the first experimental studies of spiral wave unpinning using CPEF in an excitable medium.  
However, the mechanism of how CPEF acts on a chemical wave is different from that of the cardiac excitation wave.  
In particular, we find no cut-off frequency for CPEF to unpin a chemical wave. 
We vary the pacing ratio, initial spiral
phase, and field strength. We deduced that the wave
unpins when the component of the electric field vector along  the direction of the spiral equals or exceeds a critical field strength. 
Based on this, we predict the unpinning angle as a function
of the initial position of the spiral wave, the frequency, and the strength of
the electric field. We show that our analytical formulation agrees with experimental data and
numerical results.  

\begin{figure*}[htb!]
    \centering
    \includegraphics[width=\linewidth]{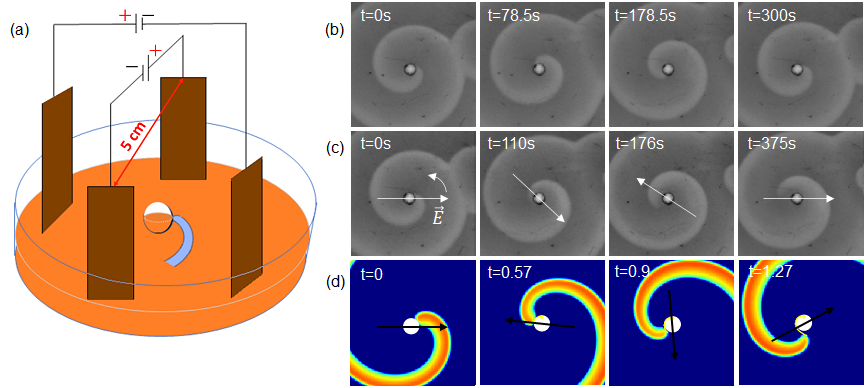}
    \caption{\textbf{(a) Schematic diagram of the experimental system:} The
	positions of two pairs of field electrodes with respect to the glass
	bead are shown schematically (not to scale).  \textbf{Unpinning of an
	anti-clockwise rotating spiral using CPEF:}  (b) An ACW 
    rotating spiral
	pinned to a spherical bead of diameter 1.2 mm in the experiment. The
	natural period of pinned spiral tip $T_{s} = 297 $ s. (c) An applied CPEF of
	strength $E_0 \simeq 1.38$ V/cm, and period $T_{E} = 125$ s unpins the spiral tip from the obstacle. 
    (d) An ACW rotating spiral pinned to an obstacle of diameter 1.0 s.u in the
	simulation with $T_{s} = 1.77$ t.u is subjected to a CPEF of strength
	$E_0 \simeq 0.6 $ and period $T_{E} =1.18$ t.u. The unpinned spiral tip drifts away
	from the obstacle at t = 1.27 t.u. The arrows show the direction of
	the applied CPEF.
    }
    \label{fig:unpinning_images}
\end{figure*}

In this paper, we focus on the unpinning of an anti-clockwise (ACW) rotating spiral using a CPEF rotating in the same direction.
We conducted our studies in the ferroin-catalyzed BZ reaction in a petri-dish, as described in detail in Ref.~\cite{Amrutha}. Briefly, we start with the following initial reagents: [$H_2SO_4$] = 0.16 M, [$NaBrO_3$] = 40 mM, [Malonic acid] = 40 mM, and [Ferroin] = 0.5 mM. The reaction mixture is embedded in 1.4 $\%$ w/v of agar gel to avoid any hydrodynamic perturbations. The single reaction layer of thickness $3 \times 10^{-3}$ m is taken in a glass petri dish of diameter $0.1$ m. 
A circular excitation wave is created at the center of the reaction medium by inserting a silver wire. By disrupting the motion of the circular wavefront, a pair of counter-rotating spirals are created. To generate a pinned spiral wave, a glass bead of diameter $1.2$ mm  is carefully placed at the tip of one of the spirals. The pinning of the spiral tip to the obstacle is confirmed after 1-2 rotations. An anticlockwise circularly polarized electric field (CPEF) is applied using two pairs of copper electrodes as in Fig.~1(a). Images of the reaction medium are recorded using a CCD camera at every $30 s$ interval for $1-2$ hours. 

To model this experiment, we use a two-dimensional Oregonator model. The model equations are given by 

\begin{equation}\label{E_uoregonator}
\frac{\partial u}{\partial t}=\frac{1}{\epsilon}(u(1-u)-\frac{fv(u-q)}{u+q})
+D_{u}\nabla^2u+M_{u}(\vec{E} \cdot \nabla u)
\end{equation}
\begin{equation}\label{E_voregonator}
\frac{\partial v}{\partial t}=u-v+D_{v}\nabla^2v+M_{v}(\vec{E} \cdot \nabla v).
\end{equation}

Here, $u$ is the activator variable, and $v$ is the inhibitor variable (corresponding to the rescaled concentrations of [HBrO2] and the catalyst, respectively). $\vec{E} = E_{0} cos(\frac{2\pi t}{T})\hat{x} + E_{0} sin(\frac{2\pi t}{T})\hat{y}$ is the circularly polarized electric field. The electric field is added as an advection term for the variables $u$ and $v$. An obstacle is added to this domain by setting the diffusion coefficient of the activator to a very low value.
Details of the model and the simulations are given in  Ref.~\cite{Amrutha}.

\begin{figure}[H]
    \centering
    \includegraphics[width=0.7\linewidth]{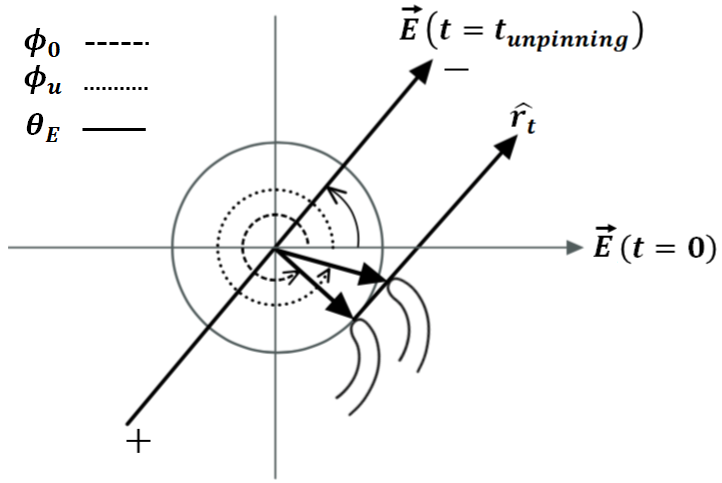}
    \caption{Schematic diagram showing the phase measurements: $\phi_{0}$ and  $\phi_{u}$ are the phase of the spiral tip at $t=0$ and at the time of unpinning respectively. $\theta_{E}$ denotes the phase of the electric field ${\vec{E}}$ and ${\hat{r}}_{t}$ is the tangential vector of spiral rotation on the obstacle boundary. All phases are measured in the anticlockwise direction from the $+x$ axis, with the obstacle center as the origin. The tail of the resultant field vector ${\vec{E}}$ marked with a $+$ sign is mentioned as the anode and the head with a $-$ sign is the cathode.
    }
    \label{fig:acw_theory}
\end{figure}

The rotating chemical wave in the BZ reaction medium can get anchored into the
boundary of the glass bead and form a very stable pinned wave, as shown in
Fig.~\ref{fig:unpinning_images}(b).  A similar situation occurs in the
numerical simulation of the model equations, where the spiral wave can get
anchored to the obstacle in the domain. It is known that this wave can be
unpinned with an electric field~\cite{Amrutha,sutthiopad2014unpinning}. 
Here we employ the circularly polarized electric field (CPEF) using two
cross-electrodes (see Fig ~\ref{fig:unpinning_images}.(a)). The CPEF can unpin the wave if the
amplitude of the electric field equals or exceeds a certain threshold value ($E_{th}$),
as shown in Fig.~\ref{fig:unpinning_images}(c).  An arrow indicates the instantaneous direction of the electric
field. Similar unpinning
is also seen in the simulations [Fig.~\ref{fig:unpinning_images}(d)]. To
understand the unpinning process, we measure the location at which the wave
unpins from the obstacle. We can quantify the spiral location by the phase of the
spiral tip on the obstacle boundary. The phase is the angle of
the spiral tip, measured in degrees from the $+x$-axis along the anticlockwise
direction with the obstacle center as the origin. The phase of the spiral when we start the CPEF is denoted by
$\phi_0$ and the phase when the spiral unpins from the boundary is denoted by
$\phi_u$ [Fig.~\ref{fig:acw_theory}]. 
The instantaneous direction of the electric
field is denoted by the angle $\theta_E$. The direction of the spiral
is along the tangent at the obstacle, and this direction is denoted by
${\hat{r}}_{t}$.   We define the pacing ratio, $p$, as the ratio of the frequency of the CPEF ($\omega_{cp}$) to that of the spiral ($\omega_s$), {i.e.,} \( p=
\omega_{cp}/\omega_{s}\). We have varied $p$ from 0.25 to 3.

Our observations can be summarised as follows: (1) The chemical wave can be unpinned with CPEF for all pacing ratios (between 0.25 to 3), provided the strength of the electric field is equal or above a threshold. 
\begin{figure}[H]
    \centering
    \includegraphics[width=\columnwidth]{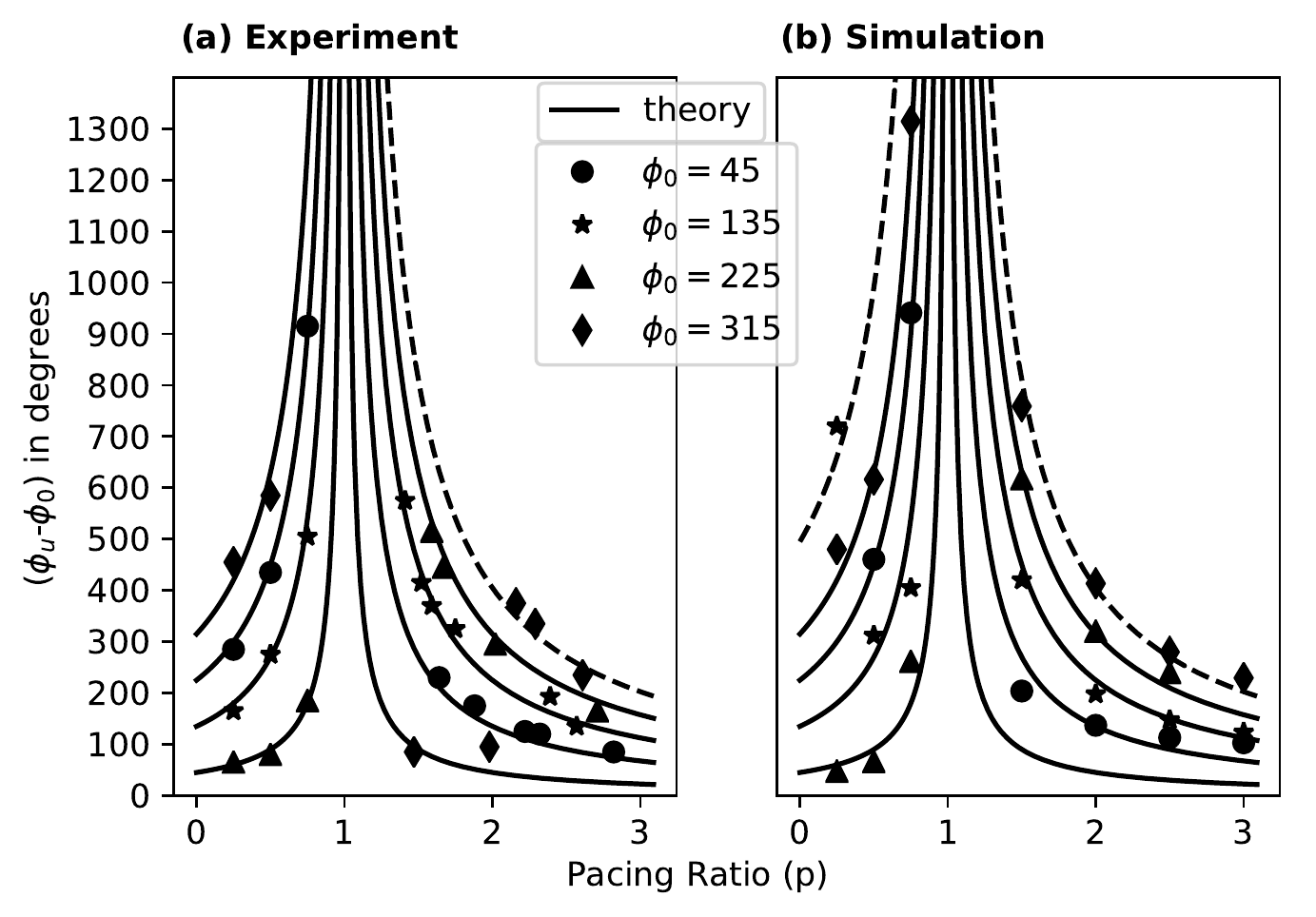}
    \caption{\textbf{Unpinning at $E = E_{th}$:} For spirals with different
	${\phi}_0$,  the phase difference (${\phi}_u$- ${\phi}_0$) is plotted (solid curve) with the pacing ratio, $p$ in (a) experiments and (b) simulations. The solid theory lines represent the phases where the unpinning condition is satisfied for the first time (Eq.\ref{eq:Eth}). The dashed lines at the top correspond to the phases where the spiral unpins when the unpinning condition is met a second time in its subsequent rotations. Most of the cases with $\phi_0 = 315^0$ show a delayed unpinning.}
   \label{fig:unpinning_EthB}
\end{figure}
There is no cut-off frequency and both overdrive pacing ($p>1$) and underdrive pacing ($p<1$) are equally effective. (2) The spiral unpinning phase ${\phi}_u$ varies linearly with $\phi_0$. It increases for overdrive pacing and decreases for underdrive pacing (Fig.~\ref{fig:unpinning_EthA}). 
(3) (${\phi}_u$- ${\phi}_0$) varies with the pacing ratio, $p$, as in Fig.~\ref{fig:unpinning_EthB}. 
(4) Unpinning is not guaranteed within one rotation of the spiral. 
In a few cases, where either the relative rotation of the spiral-field pair varies quickly (extremely overdrive or underdrive pacing), or $\phi_0$ lies close to the expected $\phi_u$ (i.e., ($\phi_u$ - $\phi_0) \approx$ 0), the spiral misses unpinning at the first expected phase.
Here, the unpinning may happen later at a phase where the unpinning condition is satisfied again (dashed lines in Fig.~\ref{fig:unpinning_EthB}).
(5) It takes several rotations for the chemical wave to unpin as $p$ approaches 1 (when the spiral and the CPEF are rotating with the same frequency). For resonant pacing ($p=1$), the wave cannot be unpinned except for a small range of initial conditions ($\phi_0$). This range increases with the strength of the electric field (Fig.~\ref{fig:Ep1}). 

\begin{figure}[H]
    \centering
    \includegraphics[width=\columnwidth]{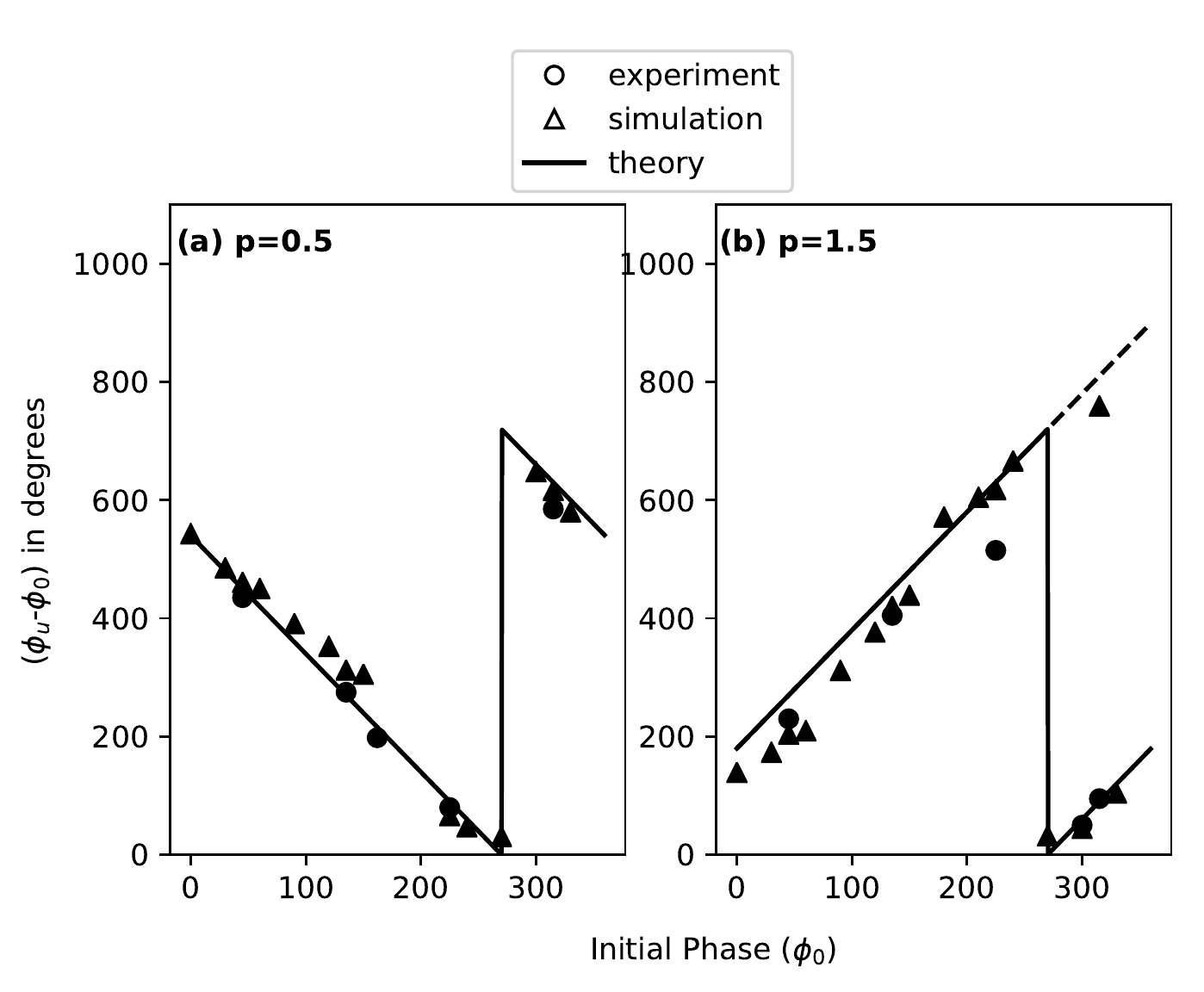}
    \caption{\textbf{Unpinning at $E = E_{th}$:} 
    (a) The spiral phase difference (${\phi}_u-{\phi}_0$) is plotted against
	${\phi}_0$ for $p$ = 0.5 (underdrive pacing). (b) same as (a) but for
	$p$ = 1.5 (overdrive pacing). In both cases (${\phi}_u-{\phi}_0$) varies linearly with ${\phi}_0$.
    The dashed line indicates the unpinning during the subsequent rotations of the electric field. To plot the theory line for ${\phi}_0 \geq 270^0$, we have added $\mp 2\pi$ to Eq.3 and Eq.4 respectively. Otherwise, the lines keep on decreasing or increasing linearly for underdrive and overdrive pacing respectively.
	Circles and triangles represent the experiment and simulation data
	respectively.}\label{fig:unpinning_EthA}
\end{figure}

These results can be analyzed in light of our recent work with the DC
electric fields~\cite{Amrutha}. We found that the electric field exerts a
retarding force on the chemical wavefront, which is maximum when the
field direction is along the direction of the wavefront. For a
CPEF with field strength $E=E_{th}$ this condition is satisfied when 
\(\vec{E}.\hat{r_t}=0\). From this, we can estimate the unpinning angle as,
\begin{eqnarray}
	\phi_u &=& \frac{p \phi_0+ 90}{p-1} ; p>1 \\
	\phi_u &=& \frac{270-p \phi_0}{1-p} ; p<1  
\label{eq:Eth}
\end{eqnarray}

When $E=E_{th}$, the wave can be unpinned only when
$\theta_E-\phi_0 = 90$. The theoretical solid curves in Figs.~\ref{fig:unpinning_EthA} and
\ref{fig:unpinning_EthB} are based on the above equation.

For a field strength
greater than $E_{th}$, the wave must unpin when the component of $\vec{E}$ along
$\hat{r_t}$ reaches the critical threshold, {\it i.e.,} $\vec{E}.\hat{r_t} \geq E_{th}$.
This condition gives an upper-bound and lower-bound for possible spiral unpinning phases
$\phi_{u}$.

For overdrive pacing with $p>1$, the unpinning phase window is given by
\begin{equation}
\frac{p \phi_0+ {\sin^{-1}}(\frac{E_{th}}{E})}{p-1}   \leq \phi_u \leq \frac{p \phi_0+\pi -{\sin^{-1}}(\frac{E_{th}}{E})}{p-1}
\label{eq:overdrive}
\end{equation}
with a width $\Delta\phi_u = \frac{\pi - 2 \sin^{-1}(\frac{E_{th}}{E})}{p-1}$. 
In most of the cases, 
the unpinning condition (Eq.~\ref{eq:overdrive}) is satisfied at the lower
bound of this range. However, unpinning is possible at any point inside the window (refer Fig.1 in the supplementary material).

For underdrive pacing i.e, for $p<1$, the unpinning phase window is 
\begin{equation}
\frac{\pi+ {\sin^{-1}}(\frac{E_{th}}{E})-p \phi_0}{1-p}   \leq \phi_u \leq \frac{2\pi-p \phi_0-{\sin^{-1}}(\frac{E_{th}}{E})}{1-p}
\label{eq:underdrive}
\end{equation}
The width of this window is $\Delta\phi_u = \frac{\pi - 2 \sin^{-1}(\frac{E_{th}}{E})}{1-p}$. Depending on the strength of the electric field, the width of the window increases. The window reduces to a point when $E=E_{th}$.

For p = 1, unpinning happens only if the following condition is satisfied.
\begin{equation}
\pi+ \sin^{-1}(\frac{E_{th}}{E})  \leq \phi_0 \leq 2\pi-\sin^{-1}(\frac{E_{th}}{E})
\label{eq:p=1}
\end{equation} 
Thus for $p=1$ the range of initial phases that lead to
unpinning increases with the field strength, $E$. Figure.~\ref{fig:Ep1}(a) shows the initial spiral
phases that lead to successful unpinning as a function of $\frac{E_{th}}{E}$.

\begin{figure}[H]
    \centering
	\includegraphics[width=\columnwidth]{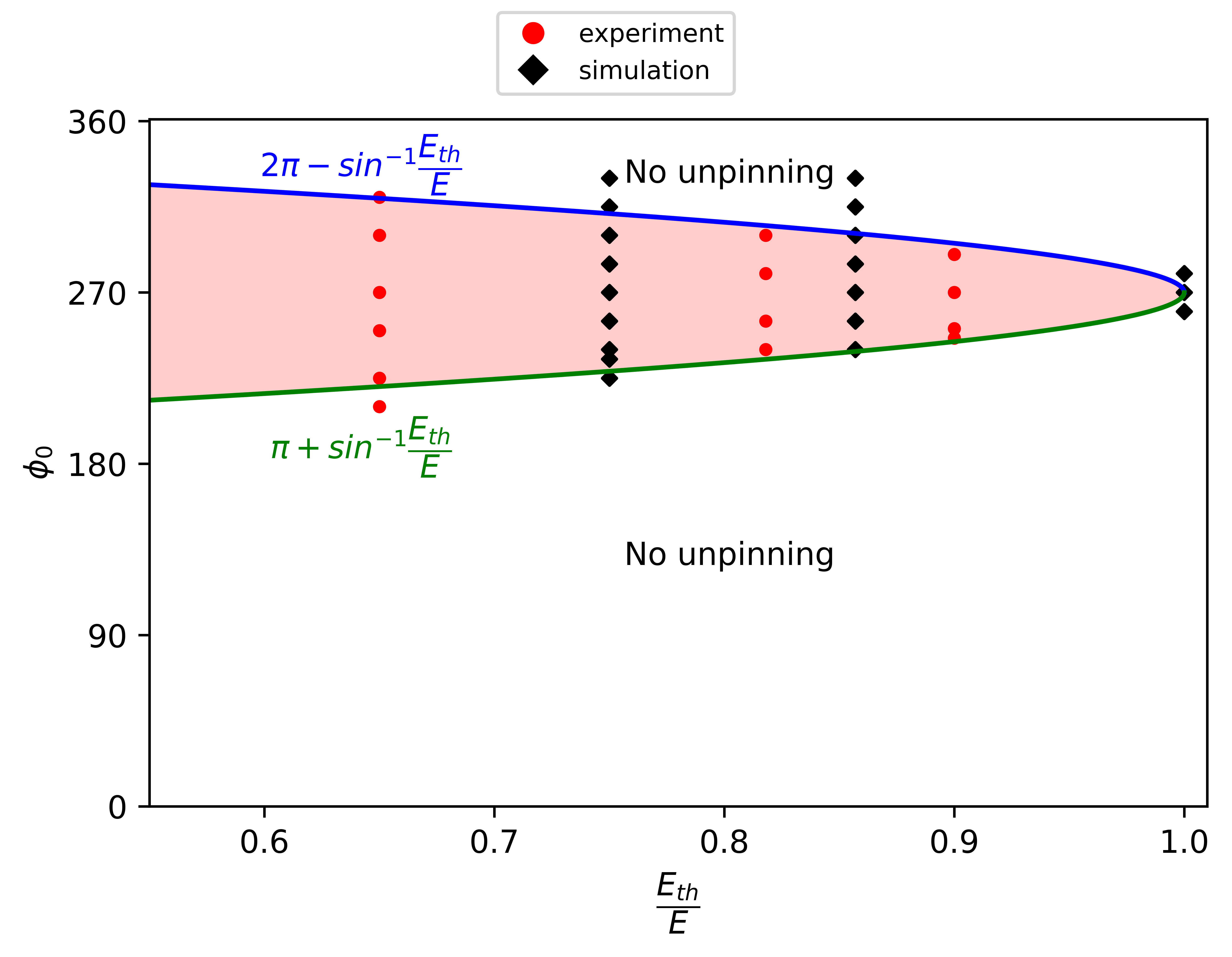}
    \caption{Unpinning of spiral wave with pacing ratio, $p$ = 1 for different
	field strength: $\pi+{\sin^{-1}}{(\frac{E_{th}}{E})}$ and
	$2\pi-{\sin^{-1}}{(\frac{E_{th}}{E})}$ are the lower and upper limit of
	the range of possible ${\phi}_0$-values which gives successful
	unpinning for $p$ = 1. The shaded region corresponds to the cases of
	successful unpinning.
    Circles and diamonds represent the experiment and simulation data
	respectively.} 
	\label{fig:Ep1}
\end{figure}

In summary, we have presented the first experimental studies using a circularly polarized electric field to unpin an excitation wave. We observed unpinning with overdrive, underdrive and resonant pacing. 
Because of the charge on the chemical wavefront, the mechanism of chemical wave unpinning differs from that in other excitable media. The wave unpins when the electric field component 
along the tangential direction of spiral propagation is equal or more than the critical
threshold; {i.e.,} when the electric force opposite to the instantaneous spiral propagation is above the threshold value. Based on this condition, we are able to predict the unpinning phase, and the same has been verified in simulations and in experiments.

The unpinning of a rotating chemical wave presents a unique physical situation.
In our studies, the unpinning happens when the anode ‘catches’ the spiral from behind while chasing it. If it fails to halt the wave and overtakes it, it has to come back again to act on it.
{i.e.,} the wave can only be unpinned while propagating away from the anode. A similar kind of asymmetry in the chemical wave behavior in an external electric field has been observed in previous studies. The speed of chemical wave propagation decreases as it propagates towards the anode and it decreases as it rotates away from it. Depending on the velocity, the size of a free spiral core varies; as the spiral accelerates, the core-size decreases, and as it decelerates the
core-size increases~\cite{agladze1992influence}. As a result, a drift of the spiral tip occurs in the medium. The drift occurs with a parallel component which is always directed towards the anode and a chirality-dependent perpendicular component~\cite{schmidt1997forced}. The phenomenon of spiral drift is addressed in numerous experimental and computational studies with dc, ac, and polarized electric fields~\cite{schmidt1997forced,chen2006drift,li2017theory}. Most of the important field effects in the BZ reaction could be explained with the electromigration of $Br^-$ and $Fe^{3+}$ ions.
Only a few studies investigated the effect of an electric field on a pinned spiral wave in the BZ reaction. In a unidirectional field, the spiral always unpins as it rotates away from the anode~\cite{sutthiopad2014unpinning,Amrutha}.
In light of previous results, we can assume that the wave can only be unpinned when it is retarded by the electric field, and not when it is accelerated by it.
During retardation, the core size increases, and the spiral can only pin weakly to obstacles smaller than the spiral core ~\cite{lim2006spiral,pumir2010wave}. This could be the reason for the asymmetric nature of the unpinning.
As a prototype model, the BZ reaction is expected to show all the qualitative features observed in other excitable systems.
On the contrary, our studies show that the chemical excitation waves interact uniquely with an external electric field.

\begin{acknowledgments}
We thank Beneesh P B, Deepu Vijayasenan, Ajith K M, K V Gangadharan, and Muhammed Mansoor C B for discussions.
Experiments were conducted using a grant (ECR/2016/000983) from Science and Engineering Research
Board, Department of Science and Technology (SERB-DST), India.

\end{acknowledgments}

\section*{Author Contributions}
S.V.A and T.K.S conceived the study. S.V.A performed the experiments, and P.S. built the experimental setup. S.P and A.S performed numerical simulations. S.V.A, A.S, and T.K.S analysed the data. A.S and T.K.S developed the theory. S.V.A and T.K.S wrote the paper. All authors helped to edit the paper.

\bibliography{Ref}

\end{document}


\newpage
\section*{Supplementary materials}
\section{Unpinning for $E > E_{th}$}
\begin{figure}[H]
    \centering
    \includegraphics{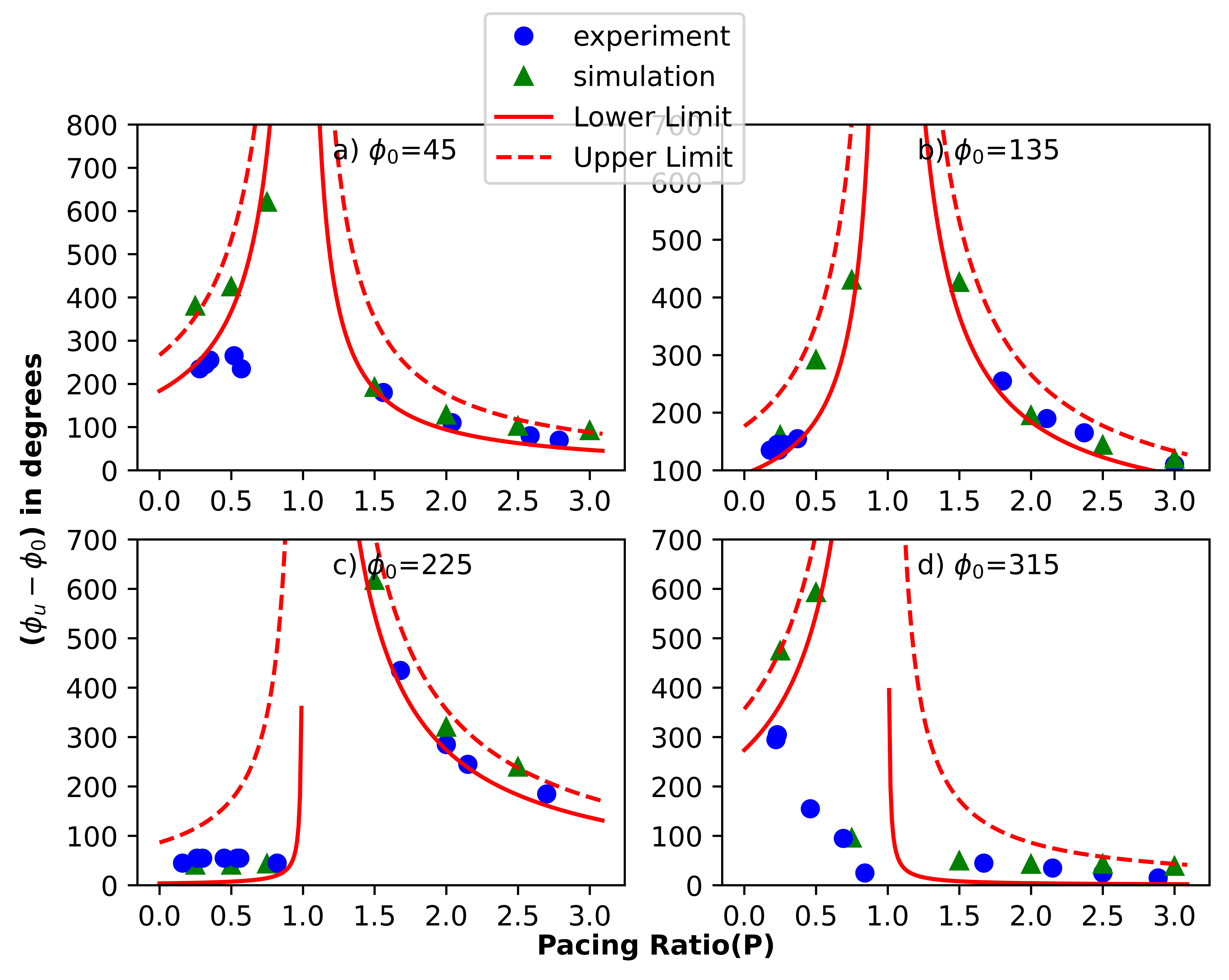}
    \caption{Unpinning at $E>E_{th}$ 
    (${\sin^{-1}}{(\frac{E_{th}}{E})}=48.95^0$): Spiral waves with different ${\phi}_0$ are unpinned in a CPEF with both under-drive ($p<1$) and over-drive pacing ($p>1$). The solid bottom line represents the lower limit of the range of possible ${\phi}_u$-values given by the relation
    ${\phi}_u=(p{\phi}_0+48.95)/(p-1)$ for over-drive pacing and ${\phi}_u=(\pi-p{\phi}_0+48.59)/(1-p)$ for under-drive pacing. The upper limit of the range of possible ${\phi}_u$-values, given by the relation ${\phi}_u=(p{\phi}_0+\pi-48.95)/(p-1)$ for over-drive pacing and ${\phi}_u=(2\pi-p{\phi}_0-48.59)/(1-p)$ for under-drive pacing, are represented by the top dashed line. For ${\phi}_0 = 315^0$, the above equations must be added with $2\pi$ to get the positive phase values.
    Circles and triangles represent the experiment and simulation data respectively. 
    }
    \label{fig:unpinning_E>Eth}
\end{figure}
Fig.\ref{fig:unpinning_E>Eth} shows the unpinning phase window at $E>E_{th}$ for different initial phases of the spiral. Here, the solid lines correspond to the lower limit, and the dashed lines correspond to the upper limit of the window according to the equations \ref{eq:overdrive} and \ref{eq:underdrive}. The unpinning always happens at a phase within this range. The width of the window varies with the field strength.

\section{Comparison between pinning obstacles of different geometry}
The results of spiral unpinning from spherical beads are presented in the paper. For comparison, we have performed similar experiments using cylindrical rods. The experimental setup is the same as in figue.\ref{fig:unpinning_images}a. A cylindrical glass rod of length $\approx$ 4 mm is inserted vertically into the medium.
\begin{figure}[H]
    \centering
    \includegraphics{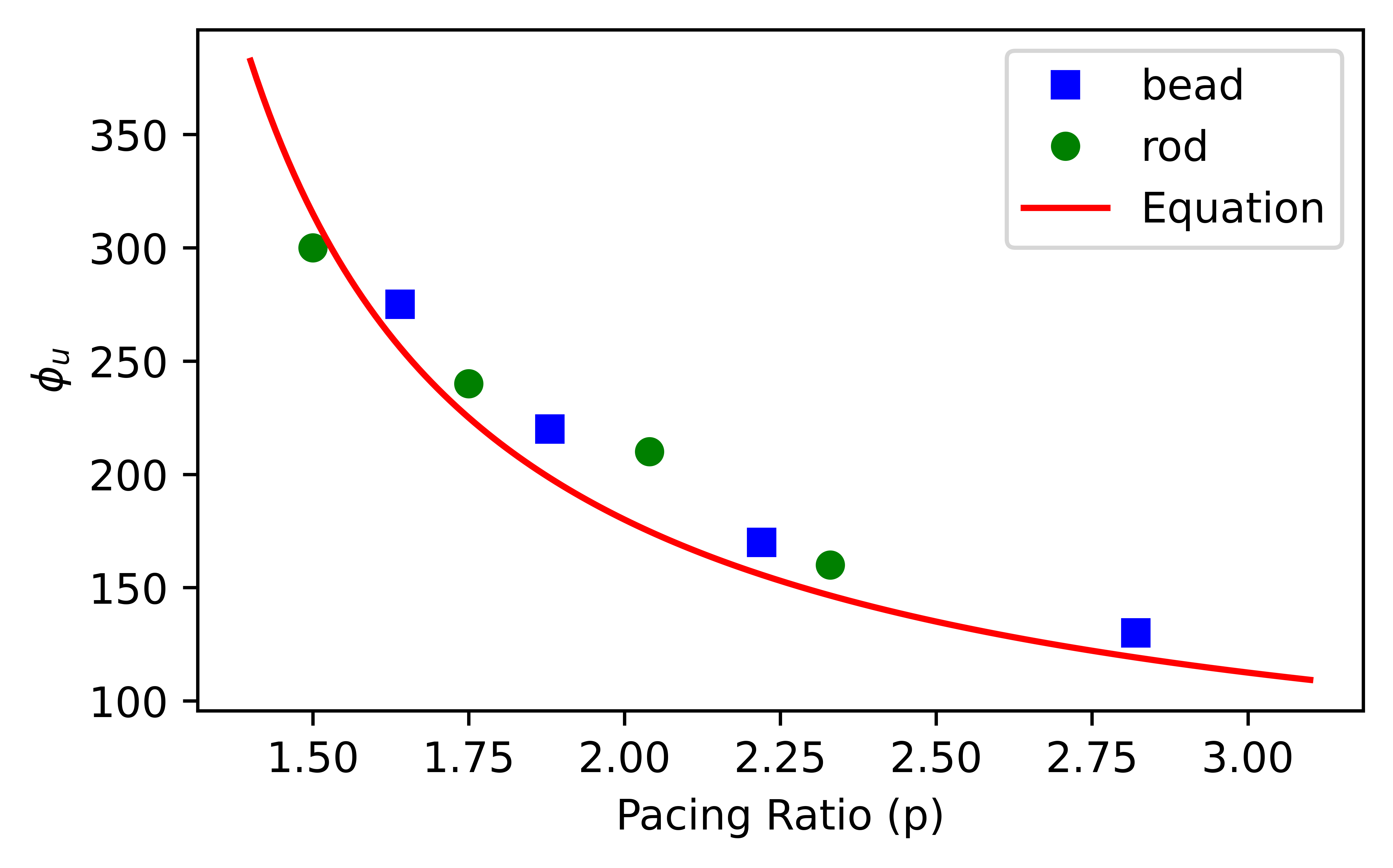}
    \caption{Comparison of unpinning of spiral pinned to spherical bead and cylindrical rod: $\phi_u$ is plotted against the pacing ratio,p where $p>1$. $\phi_0$ = $45^0$ and E = 1.38 V/cm. The diameter of the obstacles are same and equals 1.2 mm.}
    \label{fig:unpinning_comparison}
\end{figure}
Figure.\ref{fig:unpinning_comparison} shows the variation of unpinning phase with the pacing ratio for a cylindrical obstacle of radius 1.2 mm. The unpinning phases for a spherical bead have also been shown, and in both cases, unpinning occurs at phases that are consistent with the theoretical predictions.

\section{Comparison between numerical models}

In this letter, we have used a two-variable reduction of the original three-variable Oregonator model. Here we compare the unpinning studies using both two and three-variable models.

The three-variable Oregonator model consists of the following equations \cite{schmidt1997forced}. 
\begin{equation}\label{E_uoregonator3D}
\frac{\partial u}{\partial t}=\frac{1}{\epsilon}(qw-uw+u(1-u))
+D_{u}\nabla^2u
\end{equation}
\begin{equation}\label{E_voregonator3D}
\frac{\partial v}{\partial t}=u-v+D_{v}\nabla^2v+M_{v}(\vec{E} \cdot \nabla v)
\end{equation}

\begin{equation}\label{E_woregonator}
\frac{\partial w}{\partial t}=\frac{1}{\epsilon^\prime}(-qw-uw+fv)
+D_{w}\nabla^2w+M_{w}(\vec{E} \cdot \nabla w)
\end{equation}

The variables $u$, $v$ and $w$ represent the re-scaled dimensionless concentrations of $HBrO_2$, $Fe^{3+}$, and $Br^{-}$ respectively. The model parameters are $q=0.002$, $f=1.4$, $\epsilon=0.01$ as in Numerical methods in the manuscript along with an additional parameter,  $\epsilon^\prime=0.0001$. For both variables $v$ and $w$, the electric field $\vec{E}$ is added as an advection term. However, the variable $u$ is unaffected in the presence of an electric field as it corresponds to the charge-less species $HBrO_2$. The values of the ionic mobilities are $M_u$ = 0, $M_v$ = -2, and $M_v$ = 1. The simulation details can be obtained from our recent paper \cite{Amrutha}.

\begin{figure}[H]
    \centering
    \includegraphics{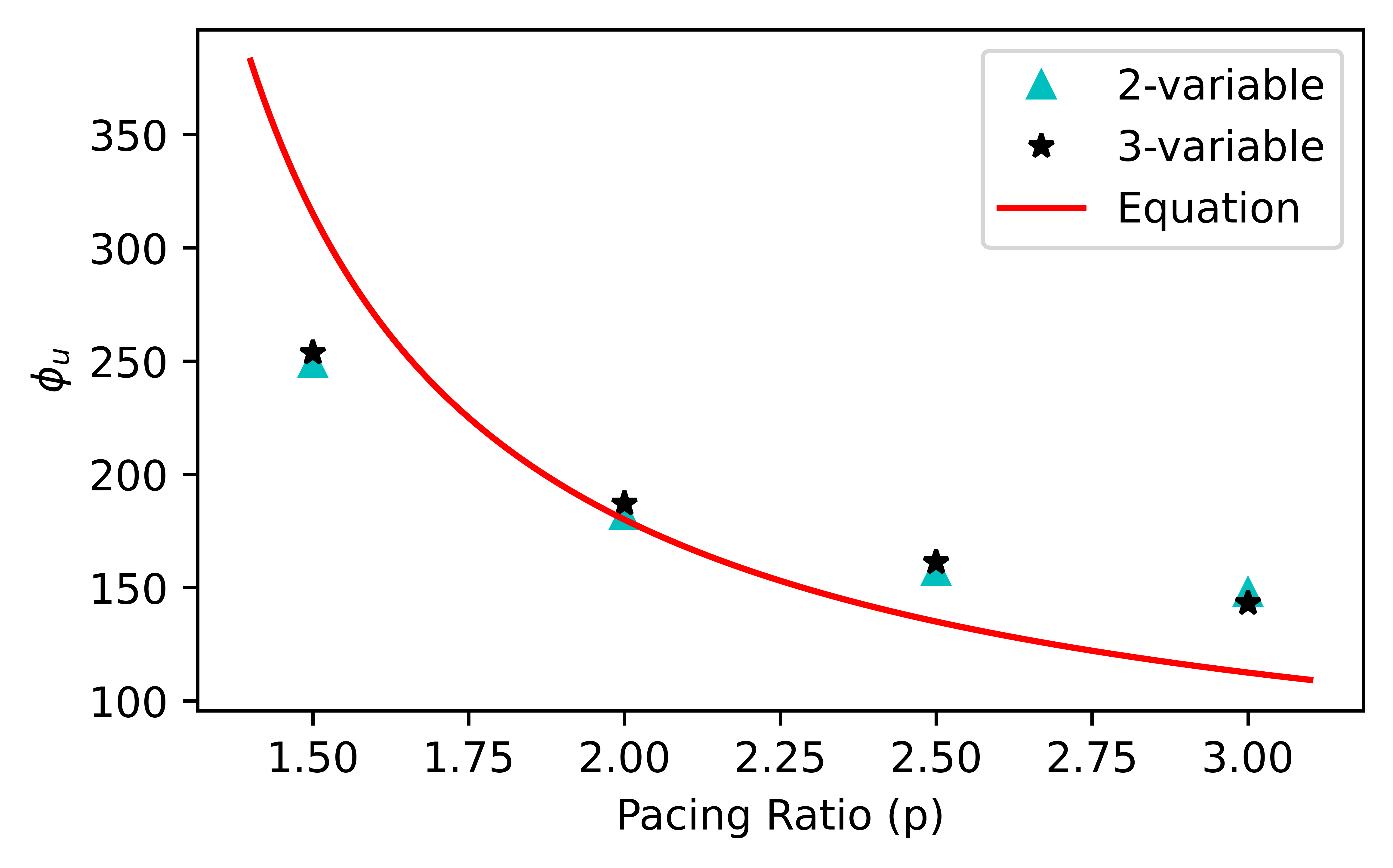}
    \caption{Comparison of spiral unpinning obtained in two and three-variable Oregonator models: $\phi_u$ is plotted against the pacing ratio,p where $p>1$. $\phi_0$ = $45^0$ and E = 0.6. The obstacle diameter is 1.0 s.u.}
    \label{fig:unpinning_comparison}
\end{figure}

Using the three-variable model, we measured the unpinning phase of an ACW spiral pinned to an obstacle of radius, r=1.0 s.u in an electric field of strength $E_{th}$ = 0.6. The unpinning is done for a fixed initial phase  with overdrive pacing. The results are in good agreement with those obtained from the two-variable Oregonator model and from the theory.

\
\bibliography{ref}